\documentclass{iau307}
\usepackage{graphicx}
\usepackage{natbib}
\usepackage{url}
\usepackage{dtklogos}
\bibpunct{(}{)}{;}{a}{}{,}

\title[Stellar Turbulence] 
{3D and Some Other Things Missing from the Theory of Massive Star Evolution}

\author[W. David Arnett]   
{W. David Arnett$^1$
 }

\affiliation{$^1$Steward Observatory\\ University of Arizona \\Tucson AZ 85721, USA\\ email: {\tt darnett@as.arizona.edu} \\[\affilskip]
}

\pubyear{2014}
\volume{307} 
\pagerange{}
\jname{New windows on massive stars: asteroseismology, interferometry, and spectropolarimetry}
\editors{G. Meynet, C. Georgy, J.H. Groh \& Ph. Stee, eds.}

\begin{document}

\maketitle

\begin{abstract}
This is a sketch of a 321D approximation which is nonlocal, and thus has nonzero fluxes of KE 
(to be published in more detail elsewhere). We plan to add this as an option to MESA.
Inclusion of KE fluxes seems to help resolve the solar abundance problem \citep{asplund}. Smaller cores may ease the explosion problems with core collapse supernova simulations.
\end{abstract}

\firstsection 
\section{Introduction}
Stars are three dimensional (3D), turbulent plasma, and much more complex than the simplified one dimensional (1D) models we use for stellar evolution. Computer power is inadequate\footnote{But perhaps close; see \cite{herwig_woodward13}.} at present for adequately resolved (i.e., turbulent) 3D simulations of whole stars.
In his review talk, \cite{casey} has illustrated this complexity and shown how it may be tamed by use of 3D simulations and Reynolds-Averaged Navier-Stokes (RANS) equations. 

A minimalist, more approximate step may be easier to implement in stellar evolutionary codes, and instructive for the closure problem.
Formally, the RANS equations are incomplete unless taken to infinite order; they must be {\em closed} by truncation at low order to be useful. This may be due to the nature of the Reynolds averaging, which allows {\em all} fluctuations rather than restriction only to {\em dynamically consistent} ones.  Closure requires additional information to remove these new, extraneous solutions. As a complement to the RANS approach, approximations which focus on dynamics are emphasized here. In the cascade, turbulent kinetic energy and momentum are concentrated in the largest eddies. We introduce a simple dynamically self-consistent model which contains the largest eddies and the Kolmogorov cascade, and examine the consequences.

Here we will concentrate on the following issues:
\begin{enumerate}
\item the turbulent cascade \citep{kolmogorov41,kolmogorov62},
\item chaos \citep{lorenz63} and time dependence \citep{arnett_meakin11},
\item boundary layers \citep{prandtl34} and surfaces of separation \citep{ll_fm_59},
\item combined mixing and burning, and
\item sensitivity of core collapse to progenitor structure
\end{enumerate}
The important and related issues of coherent treatment of pulsations, eruptions and explosions, rotation and magnetic fields, and variable mass and angular momentum must be deferred.

Erika B\"ohm-Vitense developed mixing-length theory in the 1950's \citep{vitense53,bohm-vitense58}, prior to the publication in the west of Andrey Kolmogorov's theory of the turbulent cascade \citep{kolmogorov41,kolmogorov62}. MLT might have been different had she been aware of the original work Kolmogorov had done in 1941. Edward Lorenz  showed that a simple convective roll had chaotic behavior (a strange attractor, \cite{lorenz63}). Ludwig Prandtl developed the theory of boundary layers \citep{prandtl34}. All these ideas will be relevant to our discussion, which is based upon theory and 3D simulations. 

\subsection{3D Simulations}

Numerical approaches for fluid dynamics simulations include implicit large eddy simulations (ILES) and direct numerical simulation (DNS). DNS resolves the smallest relevant scales, and procedes to larger scales until limited by computer power available. At present this limit is far smaller than stellar scales, so DNS is important in stellar problems for points of principle (e.g., \cite{pascale}). ILES includes the largest scales in the system, and procedes to smaller scales until limited by computer power available. Implict is the assumption that sub-grid scale phenomena are correctly treated. This seems to be valid for 3D turbulence using state-of-the-art methods \citep{iles}. 

In order to be useful in a 1D stellar evolution code, 3D information must be projected onto a 1D coordinate system, hence ``321D"{\footnote{The acronym ``321D" for ``projection of three  dimensions to one dimension" is due to John Lattanzio.}.
Table~\ref{tab1} gives a brief comparison of features of 
some 3D simulations which are relevant to designing 1D stellar algorithms. The simulations have different strengths and weaknesses, and tend to complement each other.

\begin{table}
\begin{center}
\caption{A Few Examples of Three-dimensional Simulations of Convection in Stars.}
\label{tab1}
\begin{tabular}{c|ccc|}\hline 
\textbf{Attribute} & \textbf{3D Atmospheres} & \textbf{Solar Convection Zone} & \textbf{Stellar Interiors} \\ 
\hline
representative & Stein \& Nordlund$^1$ & J. Toomre$^2$ & Meakin \& Arnett$^3$ \\
photosphere & Y(yes) & N(no) & N(no) \\
composition gradient & N & N & Y \\
nuclear burning & N & N & Y \\
magnetic field & Y & Y & N \\
rotation & N & Y & N \\
driving & top & top & top or bottom \\
geometry & box in star & CZ in box & box in star \\
boundary inside grid & top only & No & Yes \\
hydro & compressible & anelastic & compressible \\
\hline
\end{tabular}
\end{center}
\vspace{1mm}
\scriptsize{
{\it Notes:}\\
$^1$See \cite{stein_nordlund98}, and \cite{magic_I_14,magic_MLT_14} which also refers to other recent work.\\
$^2$See \cite{bmj_ash,brun-miesch-toomre}, and many papers in the IAUS271 proceesings \citep{juri}.\\
$^3$See  \cite{ma07b,vma13}, and \cite{arnettmv14} for an overview.\\}
\end{table}

{\bf 3D atmospheres.} 
Simulations of stellar atmospheres in 3D were pioneered by
\cite{stein_nordlund98}, see also \cite{magic_I_14} and references therein.  These simulations are one of the great successes of radiation hydrodynamics, removing the necessity for the micro-turbulence and macro-turbulent fudge factors previously used to calculate stellar spectra.
It is tempting to use MLT to fit such 3D simulations as a ``stellar engineering'' exercise to connect atmospheres to interiors, because stellar evolution codes are almost always formulated in the language of MLT. \cite{magic_MLT_14} give a clear discussion of this process, and show that MLT must be modified in at least one respect to make the identification: a ram pressure term 
must be added to MLT. It appears that such fits necessarily ignore a quantity important for stellar interiors which comes from the same term but in the energy equation. This is the flux of turbulent kinetic energy, which is incorrectly defined as zero in MLT.

These 3D stellar atmospheres generally do not contain the whole convective zone, but use a lower boundary condition which has been shown to have little effect on the predicted spectra. By the same token, this means that these simulations are not a sensitive probe of the convection at the bottom of the convection zone.


{\bf Convection, rotation and MHD.} Juri Toomre and his students and collaborators have pioneered anelastic simulations of the solar convection zone, 
focusing on global MHD and differential rotation. For 2D see \citep{hurlburt-tm2d}, and 3D \citep{brun-miesch-toomre}.
Unlike the "box in a star"  grids used in 3D atmospheres and stellar interiors, such global solutions make more serious demands on computer resources because of their larger extent, so that, other things being equal, their zoning tends to be coarser. Recent results suggest that the numerical viscosity is sufficiently low so that the flows are becoming realistically turbulent \citep{bmj_ash}.
Merging of the insights from these simulations with stellar evolution with rotation is a present challenge.

{\bf Stellar Interiors.} 
Another option for use of limited computational resourses is to examine deep interiors in which composition gradients and nuclear burning occur. By using a
 "box in a star" approach it is possible to include convective zone boundaries within the computational grid. The treatment of radiation flow is simpler (radiative diffusion).
 Due to neutrino cooling the thermal time scales become shorter, and make thermal relaxation easier to deal with; also higher luminosity as found in deep layers of red giants reduces the thermal time scale. 
Deep zones (large stratification) and pressure dilatation \citep{vma13} have been examined.
The "box in a star" approach truncates the lowest order modes, so that a natural complement is 
the pioneering work of Paul Woodward (e.g., \cite{herwig_woodward13}) to put a "whole star in  a box".

\section{A 321D Algorithm}
 This procedes in several steps:
 \begin{enumerate}
 \item add turbulent cascade,
 \item add dynamic (acceleration) equation for integral scale,
 \item balance between driving, damping, and the role of turbulent KE flux,
 \item make quantitative connections to MLT and the Lorenz model,
 \item use the steady-state Lorenz model to approximate average behavior, 
 \item use acceleration equation to define boundary behavior, and 
 \item add composition effects.
\end{enumerate}
Because an acceleration equation is used, it is straight-forward (in principle at least) to add inertial forces (centrifugal and coriolis), Lorenz forces and differential rotation.

\subsection{The Turbulent Cascade}

\cite{arnettmv14} estimate the Reynolds number to be $Re \sim 10^{18}$ at the base of the solar convection zone. Numerical simulations and laboratory experiments become turbulent for $Re \sim 10^3$, so fluid flows in stars are strongly turbulent if, as we assume for the moment, rotational and magnetic field effects may be neglected. This special, simplier case is thought to be widely but not universally appropriate to stellar interiors.

For homogeneous, isotropic, and steady-state turbulence, the Kolmogorov relation between the dissipation rate of turbulent kinetic energy, velocity, and length scale is,
\begin{equation}
\epsilon \sim v^3/\ell. \label{kolmog_eps}
\end{equation}
This is a global constraint, and applies to each length scale $\lambda$ in the turbulent cascade, so
\begin{equation}
\epsilon \sim (\Delta v_\lambda)^3/\lambda ,
\end{equation}
for all scales $\lambda$, or,
\begin{equation}
\Delta v_\lambda \sim (\epsilon \lambda)^{1 \over 3}.\label{vel}
\end{equation}
so that the velocity variation across a scale $\lambda$ is $\Delta v_\lambda$, and increases as $\lambda^{1 \over 3}$.
A description of the cascade needs both large and small scales; Eq.~\ref{vel} implies that the
largest (integral) scales have most of the KE and momentum
while the smallest have the fastest relaxation times. Simulations confirm this \citep{amy09}.

\subsection{Dynamics}
In MLT the buoyant force is approximately integrated over a mixing length to obtain an average velocity $u$ (e.g., \cite{smitha14}), 
\begin{equation}
u^2 = g \beta_T \Delta \nabla  \Big ( {\ell^2_{MLT} \over 8 H_P } \Big ). \label{mlt}
\end{equation}
Working backward, this may be expressed as
\begin{equation}
du/dt = g \beta_T \Delta \nabla - u/\tau, \label{mlt_acc}
\end{equation}
where $\ell_d \equiv  {\ell^2_{MLT} / 8 H_P }$ and $\tau = \ell_d/|u|$, and for $\Delta \nabla >0$. Multiplying by $u$ gives a kinetic energy equation,
\begin{equation}
d(u^2/2)/dt = u \cdot g \beta_T \Delta \nabla - u^2/\tau, \label{mlt_KE}
\end{equation}
for which the steady-state solution\footnote{Care must be taken with the sign of the transit time $\tau$ and the deceleration for negative $u$.} is Eq.~\ref{mlt} (only positive $\Delta \nabla$ are allowed). 

Had it been available, B\"ohm-Vitense might have identified the damping term with the Kolmogorov value (Eq.~\ref{kolmog_eps}). 
However, Kolmogorov found the damping length $\ell_d$ to be the depth of the turbulent region, so that it is not a free parameter, unlike MLT. There is a further issue; $\epsilon$ is the {\em average} dissipation rate, not the instantaneous value ($u^3/\ell_d$) which fluctuates over  time; that is, $ u \neq v$ except on average over $\tau$ (see Fig.~4 in \cite{ma07b}); we return to this below.

\subsection{Kinetic Energy Flux}
The flow is relative to the grid of the background stellar evolution code, so
\begin{equation}
d(u^2/2)/dt = \partial_t ({\bf u \cdot u})/2 + {\bf \nabla \cdot F_{KE} },
\end{equation}
where $ {\bf F_{KE} } = \rho {\bf u} ({\bf u \cdot u})/2$ is a flux of kinetic energy. The generation of the divergence of a kinetic energy flux in this way is robust for dynamic models; it occurs in the more precise RANS approach as well \citep{ma07b}. It does not occur in MLT, which assumes symmetry in velocity between upflows and downflows, a condition that is violated at even weak levels of stratification \citep{ma2010apss,vma13}. Alternatively, MLT is equivalent to Eq.~\ref{mlt_acc} with $du/dt=0$, the {\em local} approximation.

\cite{vma13,mocak14} find a {\em global}
balance between buoyant driving (at the largest scales) and turbulent damping (at the smallest). Because of the large separation of these length  scales, they are weakly coupled, giving rise to fluctuating behavior typical of turbulence. This balance (on average only)  is a new condition beyond MLT, and allows the elimination of the free parameter in MLT, the mixing length, or $\alpha=\ell/H_P$. The actual flow patterns that correspond to this global balance depend upon the details of the turbulent energy input; nuclear heating and photospheric cooling give different flows \citep{ma2010apss}, i.e., fluxes of kinetic energy.

%
%
%



We choose a vector equation for the integral scale velocity $\bf u$,
\begin{equation}
 \partial {\bf u }/ \partial t + ({\bf u \cdot }\nabla ){\bf  u} = {\bf g} \beta_T \Delta \nabla - {\bf u } / \tau.
\label{accel_eq}
 \end{equation}
The damping term is chosen to give the Kolmogorov expression for $\epsilon$, the turbulent dissipation, if Eq.~\ref{accel_eq} is dotted by $\bf u$ and averaged over a turnover time $\tau$. 

\subsection{Connections to MLT and Lorenz}
In the {\em local, steady-state,} limiting case, the left-hand side of Eq.~\ref{accel_eq} vanishes, and an equation similar to Eq.~\ref{mlt} results, but with the mixing length replaced by the turbulent damping length (essentially the lesser of the depth of the convective zone or 4 pressure scale heights, \cite{arnett_meakin11}).
With this change, {\em the cubic equation of B\"ohm-Vitense may be derived} \citep{smitha14},
and we recover a form of MLT.

If it is assumed that the integral scale motion is that of a convective roll,
{\em Eq.~\ref{accel_eq} may be reduced to the form of the classic Lorenz equations}, but with a nonlinear damping term provided by the Kolmogorov cascade \citep{arnett_meakin11}. Because of the time lag, the modified equations become even more unstable than the original ones. 

%
%
%
%
%

For 321D the stellar evolution code must be supplied a smooth time-averaged value for the convective variables; we approximate that by the steady state.
The weak coupling between large scale driving and dissipation at the small scale causes 
time dependent fluctuations of significant amplitude in luminosity and turbulent velocity;
see \cite{ma07b}, Fig.~4, for fluctuations in KE in an oxygen burning shell.
The term $ \partial {\bf u }/ \partial t$ in Eq.~\ref{accel_eq} is needed for chaotic fluctuations, wave generation and large scale dynamic behavior. A stellar evolution code must step over the shorter turnover time scales (weather) to solve for the evolutionary times (climate); this requires an average over active cells, so that there is a cellular structure in both space and time. 
The steady-state limit of the Lorenz equation gives a reasonable approximation to its average behavior, filtering out the fluctuations \citep{arnett_meakin11}; we apply the same approximation to Eq.~\ref{accel_eq} for slow stages of stellar evolution.

 
%
%

\subsection{Boundary Layers}

\begin{figure}[b]
\begin{center}
\includegraphics[width=\textwidth]{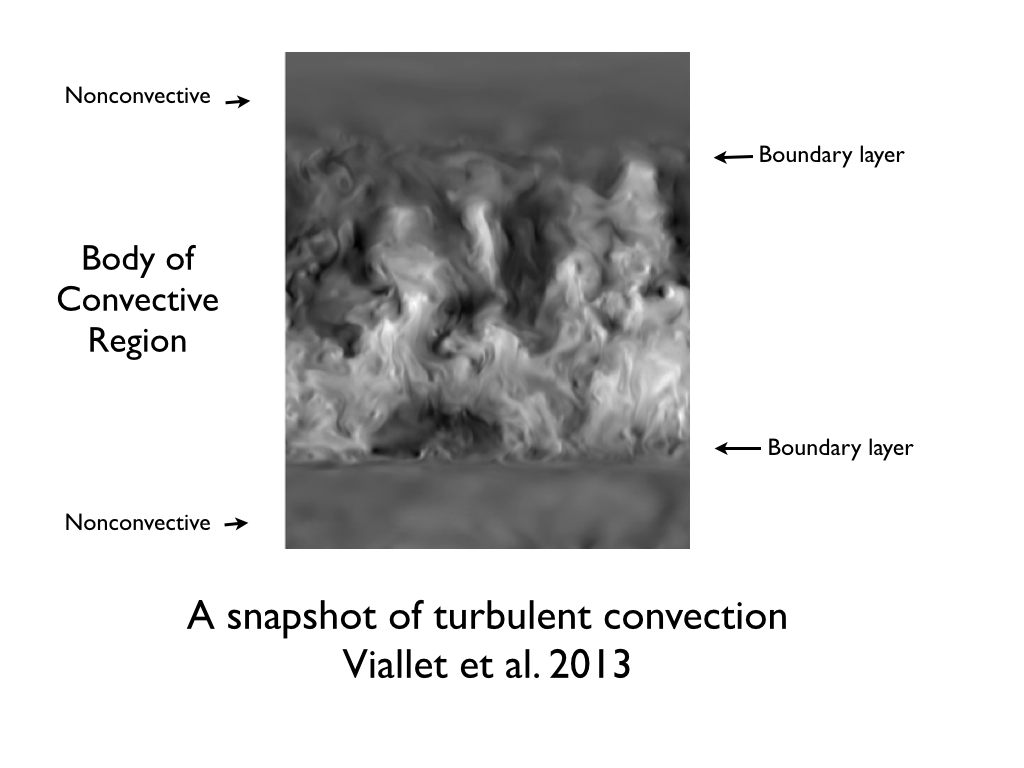} 
\caption{Boundary Layers enclosing Convection (after \cite{vma13}.}
\label{fig1}
\end{center}
\end{figure}

Fluid motion in a star may be separated into two fundamentally different flows: solenoidal flow (divergence free) and potential flow (curl free) \citep{ll_fm_59}. Fig.~\ref{fig1} shows the striking separation in flow velocities at such boundary layers (see \cite{prandtl34}. We do not claim that we have resolved these boundary layers yet\footnote{We have fewer than 10 zones across the lower boundary layer; numerical viscosity may affect the computed entrainment rate.}, but their structure and nature are important to the rate at which turbulent flow moves into or from non-turbulent regions---the entrainment rate.

Peter Eggleton  took an early step in dealing with steep gradients in composition with the introduction of an admittedly ad-hoc diffusion operator \citep{peggleton}; this numerically advantageous procedure has been widely adopted for stellar evolution.

A more physical picture results from consideration of the dynamics of the motion.
At its most elemental level, the velocity vector must turn at boundaries to maintain solenoidal flow. Most of the momentum is at the largest scales, where dissipation is least and the flow almost adiabatic, so this velocity must be the integral scale velocity. The magnitude of the acceleration required is just the centrifugal value $v^2/w$ where $w$ is the width of the turning region and $v$ the relevant velocity. Using Eq.~\ref{accel_eq} in the steady state limit, and taking $w \sim \Delta r << \ell$,
the radial component of the acceleration equation becomes
\begin{equation}
u_r \partial u_r / \partial r \sim \Delta({1 \over 2} u_r^2)/ \Delta r \sim g \beta_T \Delta \nabla .\label{turning}
\end{equation}
Sorting out the signs we see that the buoyancy must be negative for the radial kinetic energy to decrease, so that negative buoyancy decelerates the motion, and gives the required turning. 
This defines a layer at which the velocity goes to zero on average. The ``overshoot'' region has a width $w$; this material is mixed because it moves back into the convective region after it turns.

There are fluctuations, so that the layer undulates, generating waves in the neighboring stable region. In a coordinate system moving with the undulating layer,  there is a boundary  between the mixed (overshoot) region and the overlying layer which is stably stratified. 
The overshoot region has solenoidal flow which flow in the stably stratified region is potential (wave) flow. By Kelvin's theorem (conservation of circulation), entropy change (e.g., dissipation, radiative diffusion) is required to move matter from one region to the other.
Most of the turbulent momentum is in the largest scale convective motions, the integral scale. 
In simulations we see that these couple well with gravity waves\footnote{Sound waves couple well only for higher Mach number flow (\cite{ll_fm_59}, \S64).}, whose speed increases with wavelength (\cite{ll_fm_59}, \S12), so that the longer wavelengths carry most of the energy.
The shorter wavelengths are more dissipative; they are generated by nonlinear interaction of the long wavelengths (wave breaking; think water waves).
This is an example of a  mechanism for changing entropy at the boundary, allowing the turbulent region to grow. If this example is representative, such entrainment rates are not universal but depend upon local conditions at the boundary as well as turbulent velocities.
See also \cite{vma13} discussion of radiative heating at bottom of a deep (strongly stratified) convection zone.

\cite{arlette} approaches this issue from a quite different point of view, but comes to some similar conclusions.


\subsection{Mixing and Burning}
Change in composition due to nuclear burning is a fundamental feature of stellar evolution. In general stars do not have uniform composition. In stellar evolution, mixing is determined according to the criterion of Schwarzschild, or of Ledoux.
The Schwarzschild criterion is defined as ${\cal S} = \nabla - \nabla_e$,
and the Ledoux criterion is ${\cal L} = {\cal S} - \nabla_Y$, so  $\Delta \nabla = {\cal L}$ which reverts to $\cal S$ if there is no composition gradient ($\nabla_Y = 0$).

Mixing motions (solenoidal flow) result from buoyancy, pressure perturbations, or differential rotation. We focus on the buoyant acceleration, $-{\bf g} \rho'/\rho.$
Composition gradients enter the buoyancy on an equal basis with entropy gradients; 
for the simplest case of an ideal gas, $\rho'/\rho = T'/T + Y'/Y - P'/P,$ where $Y=1/\mu$ is the number of free particles per baryon, or inverse mean molecular weight.
For a more general equation of state, there are mutiplicative factors of order unity on the r.h.s.
Traditionally $P'/P$ is taken to be zero even though this is incorrect for strongly stratified cconvecton \citep{vma13}.



Brunt frequency, which is important for for asteroseismology, is
$N^2 = - {\cal L}(g \beta_T /H_P)$, indicating a fundamental connection with boundary conditions \citep{aerts}.


\section{Solar Abundances}
In the steady-state but non-local case, the $({\bf u \cdot }\nabla ){\bf  u}$ term gives  a coupling between driving regions and damping regions in the form of a flux of turbulent kinetic energy \citep{ma07b,ma2010apss}; this is assumed to  be zero by symmetry in MLT (a major flaw). {\em Even moderate stratification breaks the symmetry, and gives a finite 
flux of turbulent kinetic energy.}

Any model of the Sun which uses MLT will neglect the effect of turbulent kinetic energy fluxes.
To estimate the sign and size of the necessary changes, we scale from the simulations in \cite{vma13} for a convection zone which like the Sun is highly stratified.
The luminosity due to kinetic energy is $L_{KE} \sim -0.35 L$ near the base of the convection zone, which is significant. This negative luminosity must be balanced by increased positive radiative luminosity, to maintain a solar luminosity,  so  $L^{new} /L_\odot = 1.35 $. This may obtained by a reduction in opacity. The opacity is well known, and depends upon the composition, primarily the metal abundance. If we simply assume that the opacity scales with metalicity, $\kappa^{new}/\kappa^{old} \sim 3/4$, and the actual metalicity should shift from the stellar evolution value (from the ``standard solar model'', SSM)
to $ z^{new} \sim 0.02 (3/4) \sim 0.015$, which is the value determined from 3D atmospheres. \cite{asplund} conclude that the discrepancy between abundances determined from 3D atmospheres, and from stellar evolution  is unidentified.  The argument just given suggests that there is a significant flaw in the physics of SSM, correction for which tends to resolve the discrepancy. Until stellar models including kinetic energy flux are constructed, it seems reasonable to take the 3D atmospheric abundances as our best estimate, and attribute the disagreement to use of a local convection theory.

\section{Sensitivity to Progenitor Structure}

\begin{figure}[b]
\begin{center}
\includegraphics[width=\textwidth]{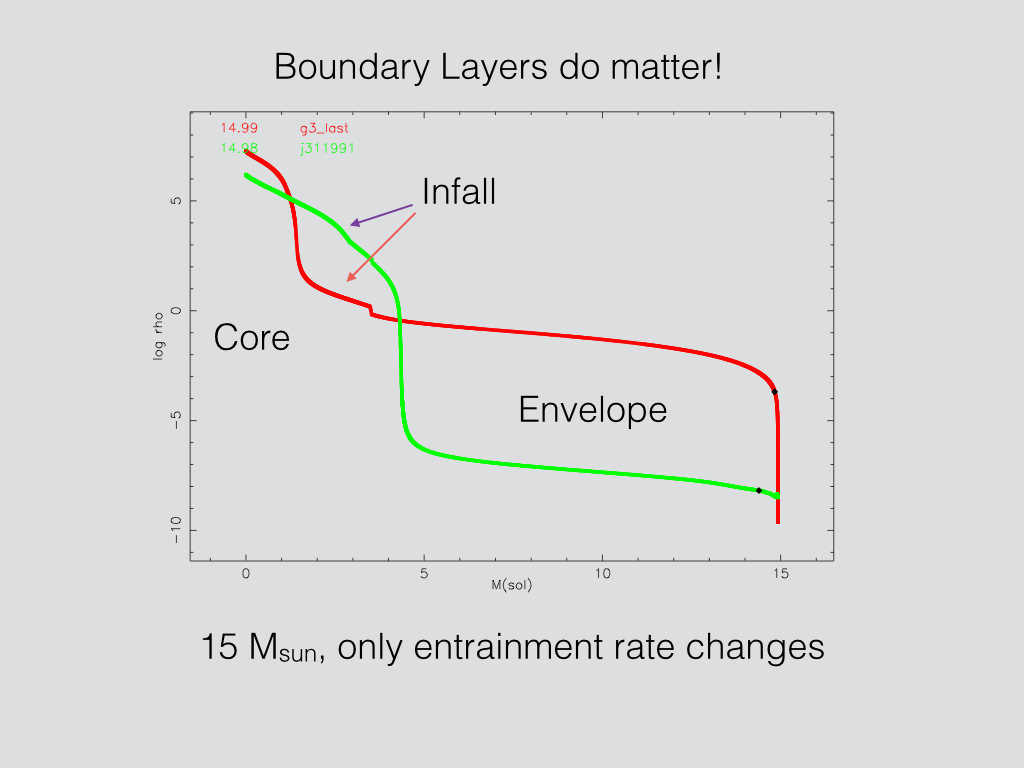} 
\caption{Changes in Density structure of $15\ M_\odot$ with Entrainment.}
\label{fig2}
\end{center}
\end{figure}

Simulations of core collapse may be sensitive to convection algorithms used to construct precollapse models. 
\cite{couch_ott13} get increased tendency toward explosion simply by adding a {\em nonradial} velocity component to the progenitor model \citep{am11}, as turbulence requires.
This also suggests changes to the initial models used by \cite{kochanek14}. 

Fig.~\ref{fig2} shows two TYCHO models at the end of oxygen burning.
The density structure is sensitive to the assumed entrainment physics. The two curves represent stars of 15$M_\odot$, with only the entrainment physics changed. Not only are the sizes of the cores changed, but the carbon/oxygen ratio is affected by entrainment as well.

Core-collapse simulations often fail because of the high rate of infall of mantle matter onto the newly formed, nuclear density core. This matter must be photo-dissociated by the explosion shock if an explosion is to occur. Lower rates of infall aid the explosion. The rate of infall is
\begin{equation}
{\dot M}_{in} = u_{in} 4 \pi r_{in}^2 \rho_{in}.
\end{equation}
The infall velocity is essentially the local sound speed in the progenitor mantle; matter falls in as a rarefaction wave moves out. To make the rate of mass infall $\dot M_{in}$ small, the initial density $\rho_{in}$ should be small.
The critical time occurs during the infall of the mantle (labeled ``Infall"); the envelope falls in too slowly to affect the explosion shock.
Clearly the curve in Fig.~\ref{fig2} with the larger core will have larger $\rho_{in}$,  and be harder to explode. This large core resulted from taking the maximum entrainment rate that was energetically allowed, throughout the evolution from the main sequence (for simplicity no mass loss or binary stripping were considered).

The small core case resulted from an entrainment rate of 0.01 of the maximum, so that the boundary dynamics was almost elastic, and consistent with analytic estimates. Smaller rates and smaller cores are possible. The highest resolution 3D simulations we have done were the most nearly elastic in the boundary layers, so we regard this as the more reasonable case. It has smaller cores that conventional progenitor models; ad-hoc diffusion smoothes gradients, leading to larger cores.

\section{Conclusions}
This is a sketch of a 321D approximation which is nonlocal, and thus has nonzero fluxes of KE 
(to be published in more detail elsewhere). We plan to add this as an option to MESA.
Inclusion of KE fluxes seems to help resolve the solar abundance problem \citep{asplund}. Smaller cores may ease the explosion problems with core collapse supernova simulations.

\bibliographystyle{iau307}
\bibliography{MyBiblio}

\begin{discussion}

\discuss{Andr\'e Maeder}{
Analytical studies suggest that transport processes by meridional circulation and by shear diffusion are strongly affected by the horizontal turbulence in differentially rotating stars. What can the 3D simulations tell us about the possible value of the horizontal turbulence?
}
\discuss{David Arnett}{Great question! We do not seem to see Jean-Paul Zahn's strongly asymmetric diffusive mixing velocities. However the overturn is effective at mixing the whole convective region, except for Si burning, which has ``cellular'' burning regions. The shellular approximation may be fine for earlier stages, up to and including oxygen burning, if we can get the boundary motion---the entrainment rates---right. The shellular approximation will break down during explosive or eruptive events due to 3D instabilities, of course.

Now suppose we spin up a non-rotating star. Turbulence is turbulence, whether driven by buoyancy or by differential rotation. At first we get slight distortions of spherical equipotentials, leading to meridional circulation, and the solenoidal flow also reacts to the inertial accelerations. Eq.~\ref{accel_eq} is a 3D vector equation, and contains these effects; the velocity includes the velocity of meridional circulation and the horizontal turbulence, at least in principle. An interesting issue is the relative importance of dissipation in the boundary layers versus in the bulk of the turbulent flow.

As the spin increases (Rossby number decreases) it will have an increasingly important influence on flow, convection, and angular momentum transport, and the importance of MHD becomes an issue (our code does not have MHD active at present). One can think of a continuous sequence from non-rotating star to accretion disk, parameterized by angular momentum, and the literature contains examples of 3D simulations all along this sequence. That said, the results are complex, and it seems that we still have much to learn.
}

\discuss{Victor Khalak}{How may stellar rotation be introduced into the equation for turbulence?}
\discuss{David Arnett}{It is already in Eq.~\ref{accel_eq} if we do the usual transformation to a rotating frame. This is a deceptively simple {\em vector} equation (the Navier-Stokes equation in the turbulent limit), which has deep connections to a lot of theoretical work on rotating stars. Andr\'e Maeder has written a beautiful development of the physics of rotating stars \citep{maeder} which starts from the Navier-Stokes equation (his Eq.~1.2).}

\discuss{Ehsan Moravveji}{In the very vicinity of the fully mixed core and the radiative layers (braking and/or entrainment layer?), what is the behavior of $\Delta \nabla$ term in your energy balance equation?}
\discuss{David Arnett}{Thanks, this is an important point that I went through too quickly. The answer may be implicit in \S2.3 and \S2.4. The $\Delta \nabla$ is a factor in the buoyant acceleration, and can change sign. It depends upon both the temperature gradient and the composition gradient. However mixing makes the composition gradient tend to zero, which changes $\Delta \nabla$ itself; this problem must be solved implicitly. In the simulations there is a transition layer in which the composition does change, but it may not yet be resolved numerically (the ``boundary layer" in Fig.~\ref{fig1}). This is the layer where the velocity field changes from solenoidal to potential flow, i.e., from convection to waves. See also the contribution by Arlette Grolsch \citep{arlette}. 

The simulations show additional complexity: the boundary is dynamic and has vigorous and fluctuating wave motion, features not in MLT. If we assume that the boundary dynamics has only a slow secular variation on average (an ``entrainment" model, \cite{ma07b}), we might use Eq.~\ref{turning} and asteroseismology to try to make progress. }

%
%
%

\end{discussion}

\end{document}